\documentclass[11pt,reqno]{amsart}

\usepackage{amsmath,amssymb,mathrsfs,amsthm}
\usepackage{graphicx,cite,cases,enumitem}
\usepackage{subfigure,epstopdf}
\usepackage[pagewise]{lineno}
\usepackage{color}
\newtheorem{theorem}{Theorem}[section]

\newtheorem{proposition}[theorem]{Proposition}

\newtheorem{defi}[theorem]{Definition}

\newcommand{\upcite}[1]{\textsuperscript{\textsuperscript{\cite{#1}}}}

\numberwithin{equation}{section}

\begin{document}

\title[Reconstruction for the periodic structure]
{Near-field reconstruction of periodic structures with superimposed illumination}

\author{Jue Wang}
\address{School of Mathematics, Hangzhou Normal University, Hangzhou, 311121, China.}
\email{wangjue@hznu.edu.cn}

\author{Yujie Wang}
\address{School of Science, Dalian Maritime University, Dalian 116026, China}
\email{wyj1028@dlmu.edu.cn}

\author{Lei Zhang$^{*}$}
\address{College of Science, Zhejiang University of Technology, Hangzhou, 310023, China}
\email{zhanglei@zjut.edu.cn}

 \author{Enxi Zheng}
 \address{School of Science, Dalian Maritime University, Dalian 116026, China}
\email{enxizheng2003@dlmu.edu.cn}

\thanks{Project supported by National Natural Science Foundation of China (No.12371420, 12271482), Zhejiang Provincial Natural Science Foundation of China (No.LY23A010004, LZ23A010006), the Scientific Research Starting Foundation (No. 2022109001429) and the Fundamental Research Foundation for the Central Universities (No.3132020168).}

\thanks{$*$ Corresponding author.}




\begin{abstract}
This paper addresses the reconstruction of periodic structures using phase or phaseless near-field measurements. We introduce a novel illumination strategy based on the quasi-periodic condition. Employing the Dirichlet-to-Neumann (DtN) map, we reformulate the scattering problem into a uniquely solvable boundary value problem. Our theoretical analysis shows the advantages of employing a fixed wavenumber superposition illumination strategy and proves that a specific wavenumber interval and fixed direction uniquely determine the grating structure. Numerical results confirm the efficacy of this illumination strategy and reconstruction algorithm, enabling super-resolution imaging from both phase and phaseless near-field measurements.
\end{abstract}

\maketitle

\section{Introduction}

Inverse scattering problems of periodic structures play an essential role in diverse areas of science and technology, such as nano-optics, diffractive optics imaging and design, optical element fabrication, and beam splitters (see, e.g.,  \cite{PR, BL2021}). In science and engineering, measuring the phase information of the wave field is usually more challenging than measuring the intensity, which is less susceptible to noise (see, e.g., \cite{K2017, JLZ2019}). Therefore, this paper aims to present a novel method for reconstructing gratings using near-field phase and phaseless data.

Several analytic and numerical techniques have been developed in recent decades to solve inverse scattering problems for periodic structures using different measurement approaches. The uniqueness of the inverse scattering problem for a lossy medium grating has been established by Bao in \cite{B1994} for the case of a single incident wave. In the context of a lossless medium grating, Kirsch \cite{K1994} demonstrated the uniqueness of an inverse grating problem using the measurements for all $\alpha$ quasi-periodic incident waves. In \cite{B-L-L}, the authors propose a continuation algorithm for reconstructing the shape of a perfectly conducting periodic structure using phaseless measurements of the scattered field with an incident plane wave. In \cite{B-Z}, the authors address a sound-soft rough surface reconstruction problem with phaseless measurements, utilizing a recursive iteration algorithm for multiple frequencies of tapered wave incidence. The transformed field expansion methods, introduced in \cite{B-L, Z-C-L-L}, offer solutions to inverse scattering problems in the near field regime (measurements taken at distances less than the incident wave's wavelength), achieving super-resolution. The authors of \cite{l_z17, l_z18} explore super-resolution imaging through subwavelength resonances and present a resonance-enhanced mode of incident wave illumination. There are noteworthy findings regarding the uniqueness of the inverse problem associated with periodic polyhedral structures \cite{B-Z-Z, EHu}. It is observed that the utilization of evanescent waves can enhance resolution and surpass the diffraction limit (see \cite{Cou}). We can see that the mathematical methods for studying inverse problems arising from different incident fields, interfaces (mediums) and observation data are quite different.

Evanescent waves were once considered a commonly used mathematical tool in wave field analysis. In 1907, Rayleigh was the first to endeavor an examination of evanescent waves by expanding the scattered field in terms of outgoing waves. Lippmann etc. further refined Rayleigh’s theory in 1954. In 1962, Twersk \cite{TV1962} emphasized that while the evanescent waves generated at the grating do not convey energy away from it, they appear to exert a substantial influence on energy redistribution. They can facilitate coupling between distinct propagating orders, thereby altering the energy distribution within the directions that yield propagating waves. Experimental demonstrations were conducted in the 1980s and 1990s, such as Near-field scanning optical microscope. Practical applications of evanescent waves can be categorized into two groups: first, cases where the energy linked with the wave is employed to stimulate other phenomena, exemplified by the total internal reflection fluorescence microscope; and second, instances where the evanescent wave facilitates coupling between two media permitting traveling waves, as observed in wave-mechanical tunneling, generally referred to as evanescent wave coupling. In 1993, Bryngdahl \cite{BO1993} pointed out that “evanescent waves cannot be used for transferring optical information over any appreciable distance. Therefore, they always have to be used in combination with homogeneous propagating waves in such a way that propagating waves are converted into evanescent waves and/or vice versa”.
Research on the generation of evanescent waves and their applications in imaging has made significant progress. However, there have been no complete answers to the fundamental questions regarding the origin of near-field optical interactions until now \cite{MO2020}.

Our work draws inspiration from the above idea of using the incident wave to control scattering waves. By controlling the incident field, the observation data at the same position can be used to solve the inverse problem more efficiently. However, the incident field needs to meet the quasi-periodic condition for the periodic structure's scattering problem. Unfortunately, the plane wave superposition usually does not meet this condition. The study of a superposition illumination strategy for the periodic structure reconstruction from near-field data had little been reported to our best knowledge.
Hence, the majority of our efforts will be dedicated to presenting an effective method with a superposition illumination strategy for solving the super-resolution imaging problem of a grating from phase or phaseless measurements, as well as demonstrating that the reconstruction results surpass those obtained with a single incidence.

The paper is organized as follows. Section \ref{section: problem}, We introduce a superposition illumination wave and formulate the scattering problem in periodic structures. By an exact transparent boundary condition (TBC), we present a Dirichlet boundary value problem for the Helmholtz equations. In Section \ref{section:ip}, it provides a theoretical analysis of the advantages of employing a fixed wavenumber superposition illumination strategy and obtains the uniqueness of the inverse problem for multi-frequency incidence. Furthermore, the iterative algorithms for reconstructing the periodic structure by the near-field measurements are obtained. We showed some numerical results in Section \ref{section: Numerical} to illuminate the proposed method's effectiveness. Finally, we summarize the paper in Section \ref{section: Con}.

\section{Scattering problem}\label{section: problem}

\begin{figure}
\centering
\includegraphics[width=0.50\columnwidth]{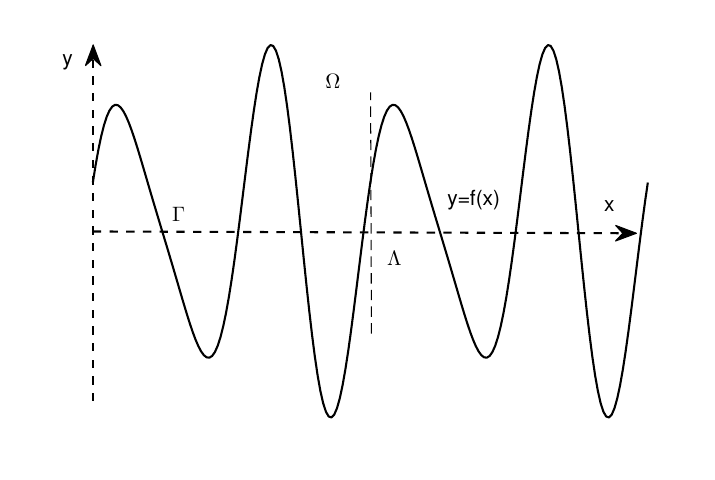}
\caption{The problem geometry of a one-dimensional grating.}
\label{bg}
\end{figure}

We consider a grating satisfies the periodic condition in the $x$ direction, as shown in Fig. \ref{bg}. Suppose that the grating
surface is described by a function $y=f(x)$
which is periodic with period $\Lambda$, i.e.,
$$
f(x+n\Lambda)=f(x),\quad\quad \quad n\in\mathbb{Z}.
$$

Let $C_{per}^{2}(0,\Lambda):=\{g\in C^{2}(0,\Lambda):~g(0)=g(\Lambda)\}$,
then we have $f\in C_{per}^{2}(0,\Lambda)$. Define
\[
\Omega_{f}=\{(x, y)\in\mathbb{R}^2: ~0< x< \Lambda,~y>f(x)\},\quad\quad
\Omega_{h}=\{(x, y)\in\mathbb{R}^2: ~0< x< \Lambda,\ f(x)<y<h\},
\]
and
\[
\Gamma:=\{(x, y)\in\mathbb{R}^2:~0< x< \Lambda,~ y=f(x)\}, \quad
\Gamma_{h}=\{(x, y)\in\mathbb{R}^2: ~0< x < \Lambda,~y=h\},
\]
where
$h>\max\limits_{0<x< \Lambda} f(x)$.

The grating is illuminated from above by a superposition incident strategy
\begin{equation}\label{IF}
u^{\rm I}(x,y) := u^{\rm inc}(x,y)+\widetilde{u^{\rm inc}}(x,y),
\end{equation}
where
\begin{equation}\label{Ipw}
u^{\rm inc}(x,y)=e^{{\rm i}\alpha x-{\rm i}\beta y},
\end{equation} 
and
\begin{equation}\label{Iew}
\widetilde{u^{\rm inc}}(x,y)=e^{{\rm i}\widetilde{\alpha} x+{\rm i}\widetilde{\beta}|y-H|},\quad\quad H>h.
\end{equation} 
Here $\alpha=\kappa \sin\theta,\beta=\kappa \cos\theta \ (|\theta|<\frac{\pi}{2})$ and $\kappa$ is a positive constant wavenumber.
Let
$$
\widetilde{\alpha}=\alpha+\frac{2\pi}{\Lambda}n_0,
$$
then there exists $n_0\in\mathbb{Z}^+$ such that
$$
\widetilde{\alpha}^2+\widetilde{\beta}^2=\kappa^2<\widetilde{\alpha}^2,
$$
which implies that $\widetilde{\beta}={\rm i}\sqrt{\widetilde{\alpha}^2-\kappa^2}$ is a pure imaginary number.
Obviously, the incident field
\[
u^{\rm I}(x+\Lambda,y)= u^{\rm inc}(x+\Lambda,y)+\widetilde{u^{\rm inc}}(x+\Lambda,y)=u^{\rm I}(x,y)e^{{\rm i}\alpha \Lambda}
\]
is quasi-periodic with $\alpha$, and satisfies
\begin{equation}\label{IFeq}
\Delta u^{\rm I}(x,y)+\kappa^2 u^{\rm I}(x,y) =0\quad\text{in} ~ \Omega_{f}.
\end{equation}
When the incident field $u^{\rm I}(x,y)$ scatters, giving rise to a scattered field $u^s(x,y)$ that satisfies
\begin{equation}\label{SF}
\Delta u^s(x,y)+\kappa^2 u^s(x,y) =0\quad\text{in} ~ \Omega_{f},
\end{equation}
and the boundary condition
\begin{equation}\label{DBC}
u^s(x,y)+u^{\rm I}(x,y) = u(x,y)=0\quad\text{on} ~ \partial\Omega_{f},
\end{equation}
where $\partial\Omega_{f}=\{(x,y)\in\mathbb{R}^2: ~0< x< \Lambda,~y=f(x)\}$.

Motivated by uniqueness, we seek the $\alpha$-quasi-periodic solution $u^s(x,y)$,  which has Rayleigh's expansion
\begin{equation}\label{RE}
u^{s}(x,y)=\sum_{n\in\mathbb{Z}} u^{s}_{(n)}(h) \exp({\rm i}\alpha_n x+ {\rm i}\beta_n (y-h)), \quad \quad y>h,
\end{equation}
where $\alpha_n=\alpha+\frac{2\pi}{\Lambda}n$,
\[
\beta_n=\left\{
\begin{array}{lll}
(\kappa^2-\alpha_n^2)^{1/2} &\text{for} ~\kappa\geq|\alpha_n|,\\[5pt]
{\rm i}(\alpha_n^2-\kappa^2)^{1/2} &\text{for} ~\kappa<|\alpha_n|,
\end{array}
\right.
\]
and
\[
u^{s}_{(n)}(h):=\frac{1}{\Lambda}
\int_{0}^{\Lambda}u^{s}(x,h)\exp(-{\rm i}\alpha_n x) d x
\]
is the Fourier coefficient. Assume $\beta_n \neq 0$ to exclude possible resonance, and $E=\{n\in \mathbb{Z} : |\alpha_n| > k \}$,  obviously,
\begin{equation}\label{RE1}
u^{s}(x,y)=\sum_{n\in\mathbb{Z} \backslash E} u^{s}_{(n)}(h) \exp({\rm i}\alpha_n x+ {\rm i}\sqrt{\kappa^2-\alpha_n^2} (y-h))+\sum_{n\in E} u^{s}_{(n)}(h) \exp({\rm i}\alpha_n x-\sqrt{\alpha_n^2-\kappa^2}  (y-h)), \  y>h,
\end{equation}
for $n \in E$ is an evanescent wave, each
term for $n\in\mathbb{Z} \backslash E$ is a
propagating wave, i.e., the $n$th order diffracted wave. It is worth noting that the above expansion does not hold in the grating grooves.

Furthermore, the  problem  \eqref{SF}-\eqref{RE} admits a variational formulation in a bounded periodic cell in $\mathbb R^2$, enforcing a transparent boundary condition on $\Gamma_h$
\begin{equation}\label{DTN}
\partial_{y} u=Tu+\rho \quad\quad\quad \text{on}  ~ \Gamma_{h},
\end{equation}
where
\begin{equation*}
Tu=\sum_{n\in\mathbb{Z}} {\rm i}\beta_n u_{(n)}(h)
\exp({\rm i}\alpha_n x) ,\quad\quad\rho=\partial_{y} u^{\rm I}-Tu^{\rm I} \ \ \text{on}~\Gamma_{h}.
\end{equation*}

Hence, we propose the following boundary value problem: find a quasi-periodic function $u$ such that
\begin{equation}\label{BVP}
\left\{
\begin{array}{lll}
\Delta u + \kappa^2 u=0 & ~ \text{in} ~ \Omega_{h},  \\[5pt]
u=0 & ~ \text{on}  ~ \Gamma, \\[5pt]
\partial_{y} u=Tu+\rho  & ~ \text{on}  ~ \Gamma_{h}.
\end{array}
\right.
\end{equation}

Given a superposition incident field $u^{\rm I}$, which is consistent with the results of the superposition interaction of evanescent waves and corresponding certain propagating orders Raleigh waves described in \cite{BO1993}. The \emph{direct problem} is to solve the problem \eqref{BVP}.
Obviously, the operator $T$ is continuous from $H^s_{qp}(\Gamma)$
to $H^{s-1}_{qp}(\Gamma)$ for any $s\in \mathbb{R}$,
where $H^s_{qp}(\Gamma)$ stands for the quasi-periodic trace functional space.
Let
$$
H^s_{qp}(\Omega_h)=\{v\in H^s(\Omega_h):~v(0,y)=v(\Lambda,y)e^{-{\rm i}\alpha\Lambda},\ v|_{\Gamma}=0\}
$$
denote the Sobolev space of functions on $\Omega_h$, which are quasi-periodic in $x$.
The existence and uniqueness of quasi-periodic solutions in $H^s_{qp}(\Omega_h)$ can be established along the lines of the proof presented in Chapter 3 of \cite{BL2021}.

\begin{proposition}
For $f\in C_{per}^{2}(0,\Lambda)$, 
the boundary value problem \eqref{BVP} possesses a unique solution $u\in H^1_{qp}(\Omega_h).$
\end{proposition}

\section{Inverse scattering problem}\label{section:ip}

In this section, we address the reconstructing a periodic curve subjected to Dirichlet boundary conditions.  The \emph{inverse problem} is to recover the periodic structure using near-field data obtained from either phased or phaseless measurements, which correspond to a superposition of incident fields.


First, define the following layered region
\[
D=\{(x, y)\in\mathbb{R}^2: \ f(x)<y<g(x)\},
\]
where $f,g\in C_{per}^{2}(0,\Lambda)$ denote two $\Lambda$-periodic functions with $f(x) <g(x)$.
Note that we study quasi-periodic solutions for the Helmholtz equation with momentum $\alpha=\kappa \sin\theta$.
Hence, we consider the following definition.

\begin{defi}
Let $\theta\in (-\frac{\pi}{2}, \frac{\pi}{2})$ be fixed. The positive constant wavenumber $\kappa$ is called a Dirichlet eigenvalue of the layer $D$ if there exists a $\alpha$-quasi-periodic nontrivial solution
$v \in C^2(D) \cap C(\overline D) $ with momentum $\alpha=\kappa \sin\theta$ of the Helmholtz equation
$$
\Delta v+ \kappa^2 v=0 \quad\quad\text{in}~D
$$
with boundary condition $v = 0$ on $\partial D$. The function $v$ is called an eigenfunction.
\end{defi}

In the case of bounded domains, it is well known that the eigenvalues of $-\Delta$ form a discrete set. Furthermore, in the case of a periodic layered region $D$, it follows from
the set of eigenvalues of $-\Delta$ in $D$ is discrete.
The preceding result constitutes a foundation for proving the uniqueness of the inverse problem in the multi-frequency setting. Let $\tilde{u}$ be a solution of the boundary value problem \eqref{BVP}
with $\Gamma$ replaced by
$$
\widetilde{\Gamma}:=\{(x, y)\in\mathbb{R}^2:~0< x < \Lambda,~ y=\tilde{f}(x)\},
$$
where $\tilde{f}\in C_{per}^{2}(0,\Lambda)$,
but for the same superposition incident field $u^{\rm I}$ satisfying \eqref{IF}.

\subsection{Inverse problem with a superposition illumination strategy}

Why illuminate with a superposition wave? Consider a simple scattering problem with a planar interface. Significant differences arise between the scattered fields produced by a superposition wave and a plane wave. The translational invariance observed in the specular reflection of a time-harmonic plane wave at different positions is a consequence of both the linear nature of the wave equation and the properties of specular reflection. In contrast, the scattered field produced by a superposition incident field $u^{\rm I}$ which contains information about evanescent waves. Fortunately, these challenges in measuring wave fields can be circumvented.

Analyzing the wave field within the context of grating scattering presents significant challenges, particularly regarding internal wave propagation within the grating grooves. Solely relying on phased boundary measurements obtained from a single incident plane wave \eqref{Ipw} is demonstrably inadequate for reconstructing the complete shape of a grating. This is in contrast to the established limitations, where multiple examples demonstrate the impossibility of determining the minimum period of a periodic structure solely based on a boundary measurement corresponding to a single incident plane wave \cite{X-Z-Z}. Building upon the spectral theory  that established in \cite{FK1997} for multi-frequency plane wave incidence, we show that the uniqueness can also be established for the case of continuous multi-frequency superposition incident waves.

Set $\theta\in (-\frac{\pi}{2}, \frac{\pi}{2})$ be fixed and
$f,\tilde{f}\in C_{per}^2(0,\Lambda)$ be $\Lambda$-periodic functions, and
$$
H>h > \max\{f(x),\tilde{f}(x)\}.
$$ For the incident field $u^{\rm I}$ satisfying \eqref{IF}, if we have
\begin{equation}\label{Measure1}
u(x,h)=\tilde{u}(x,h),
\end{equation}
then from \eqref{RE1} and \eqref{Measure1}, we conclude that $\tilde{u}=u$ in the region $\{(x,y)\in \mathbb{R}^2 : ~0< x < \Lambda,~ h\leq y<H\}$  and thus, by unique continuation,
$u$ and $\tilde{u}$ coincide in the region $\{(x,y)\in \mathbb{R}^2 : ~0< x < \Lambda,~ H \geq y \geq\max\{\tilde{f}(x),f(x)\}\}$.
We assume $f\neq\tilde{f}$
and distinguish between two cases.

\begin{figure}[htbp]
\centering
\subfigure[]{
\includegraphics[width=6.5cm]{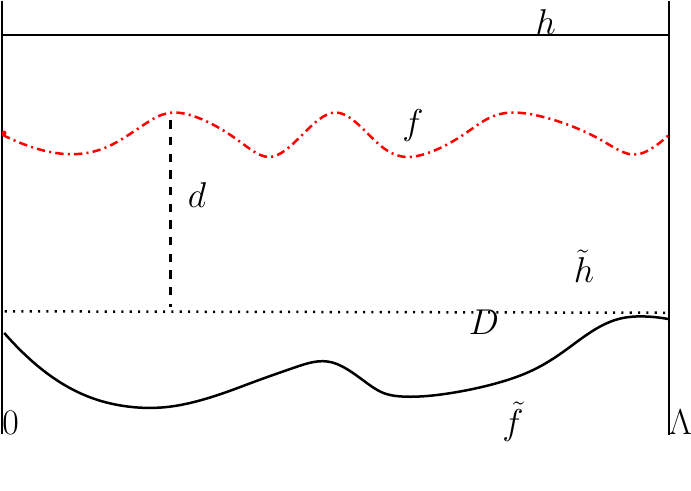}
}
\quad
\subfigure[]{
\includegraphics[width=6.5cm]{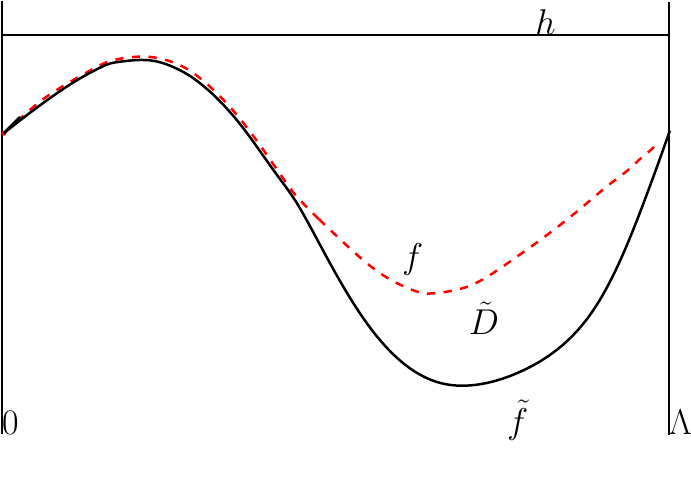}
}
\caption{Schematic of the relative positions of the two gratings}
\label{SY1}
\end{figure}

{\bf Case 1:} Without loss of generality, we assume that $\tilde{f}(x) < f(x)\in C_{p}^2(0,\Lambda)$ as shown in Figure \ref{SY1}(a).
The wave field 
$$
\tilde{u}(x,f(x))=u(x,f(x))=0,\ x \in (0,\Lambda),
$$
and
$$ 
\tilde{u}(x,\tilde{f}(x))=0,\ x \in (0,\Lambda),
$$ 
they are satisfy the quasi-periodic conditions. In the layer $D:=\{(x,y)\in\mathbb{R}^2:0<x<\Lambda,\ \tilde{f} (x)<y<f(x)\}$,
$\tilde{u}$ is a Dirichlet eigenfunction of the negative Laplacian in $D$ with $\kappa^2$.  However, the set of eigenvalues of $-\Delta$ in $D$ is finite set (see e.g., \cite{GT1977, FK1997}). There is a $\kappa\in[\kappa_{min},\kappa_{max}]$ such that the $\tilde{u}=0$ in $D$, and by unique continuation, $\tilde{u}=0$ in $\Omega_{\tilde{f}}$. 

{\bf Case 2:} There exists $a_0,a_1\in [0,\Lambda]$ with $\tilde{f} (a_j) = f(a_j),\ (j=0,1, a_0<a_1)$ and, without loss of generality, $\tilde{f} (x) <f(x)$ for all $x\in (a_0, a_1)$. Within the bounded domain $\tilde{D}=\{(x,y)\in\mathbb{R}^2:~a_0<x<a_1,\ \tilde{f} (x)<y<f(x)\}$, depicted schematically in Figure \ref{SY1}(b), we can obtain that $\Delta \tilde{u} + \kappa^2 \tilde{u}=0 \ \text{in} \ D$,

\begin{equation}\label{BVPD}
\left\{
\begin{array}{lll}
\Delta \tilde{u} + \kappa^2 \tilde{u}=0 & ~ \text{in} ~ \tilde{D},  \\[5pt]
\tilde{u}=0 & ~ \text{on}  ~ \partial \tilde{D}=S\cup\tilde{S},
\end{array}
\right.
\end{equation}
The domain $\tilde{D}$ may only be inside the grating groove. The same arguments as in Case 1 yield the existence of a quasi-periodic solution $\tilde{u}$ for every $\kappa\in[\kappa_{min},\kappa_{max}]$. This contradicts that the set of eigenvalues is discrete. So finally we have $f=\tilde{f}$.

We obtain the following results regarding the uniqueness of the inverse problem.
\begin{proposition}\label{PropE}
Let $\theta\in (-\frac{\pi}{2}, \frac{\pi}{2})$ be fixed and
$f,\tilde{f}\in C_{per}^2(0,\Lambda)$ be $2\pi$-periodic functions, where
$h > \max\{f(x),\tilde{f}(x):~x\in (0,\Lambda)\}$.
For all superposition incident fields $u^{\rm I}$ satisfying \eqref{IF} corresponding to wavenumbers $\kappa\in[\kappa_{min},\kappa_{max}]$ for some $\kappa_{max}>\kappa_{min}>0$, if we have
$u(x,h)=\tilde{u}(x,h)$ on $\Gamma_h$, then $f=\tilde{f}$.
\end{proposition}
The minimum number of incident waves required to generate measurements that uniquely determine the grating remains an open question. However, the superposition of such incident waves offers distinct advantages for numerical reconstructions. Ill-posedness poses a challenge when directly employing noisy evanescent wave measurements, leading to substantial errors in inversion results. Therefore, obtaining near-field data with controlled noise levels is crucial. The superposition strategy, denoted by  $u^{\rm I}$ addresses these concerns by enhancing the signal-to-noise ratio of measurements while simultaneously resolving the issue of up-down translational invariance and facilitating super-resolution imaging. These benefits will be further demonstrated in the subsequent numerical results.

\subsection{Reconstruction algorithm}

In this subsection, we propose an algorithm for reconstructing the periodic structure by the near-field measurements and phaseless near field measurements respectively. Note that, for $2\pi$-periodic function $f\in C_{per}^2(0,2\pi)$, there is the Fourier series expansion as follows:
\begin{equation}\label{FSE}
f(x)=\gamma_{0}+\sum_{k=1}^{\infty}\big(\gamma_{2k-1}\cos(kx)+\gamma_{2k}\sin(kx)\big),
\end{equation}
where
$$
\gamma_{0}=\frac{1}{2\pi}\int_{-\pi}^{\pi} f(x){\rm d}x,\quad
  \gamma_{2k-1}= \frac{1}{\pi}\int_{-\pi}^{\pi} f(x)\cos(kx){\rm d}x,\quad
  \gamma_{2k}=\frac{1}{\pi}\int_{-\pi}^{\pi} f(x)\sin(kx){\rm d}x.
$$

For reconstructing an approximated grating profile, we think it is natural to start with the finite sum in \eqref{FSE} as
\[
f(x)\simeq \gamma_{0}+\sum_{k=1}^{M}\big(\gamma_{2k-1}\cos(kx)+\gamma_{2k}\sin(kx)\big),
\]
and the problem of reconstructing the the Fourier coefficients from
a knowledge of the near-field measurements. In the context of general periodic function reconstruction, determining the optimal number of truncated terms within the Fourier expansion remains a crucial issue. This parameter plays a crucial role in regularization, analogous to the function of spectral truncation regularization.

Building upon the preceding results, we now discuss the reconstruction algorithm to different scenarios\upcite{ZWFL, LZ}. We first consider the case of phase data, which is, find $\widehat{\gamma}^{*}$ such that
\begin{equation}\label{PD1}
\int_{0}^{2\pi}\big|\mathcal{P}(\widehat{\gamma}^{*})
-u^{mes}\big|^{2}dx= \min_{\widehat{\gamma}\in \mathbb{R}^{2M+1}}\bigg(\int_{0}^{2\pi}\big|\mathcal{P}(\widehat{\gamma})
-u^{mes}\big|^{2}dx\bigg).
\end{equation}
where
\begin{equation}\label{OP}
\mathcal{P}(\widehat{\gamma}):=u^{s}_{\widehat{\gamma}}(x,h),
\end{equation}
and $\widehat{\gamma}=[\gamma_0,\gamma_1,\cdots,\gamma_{2M}]^{\top}$.
We define a discrete form of the objective function as
\begin{equation}\label{PD2}
C(\widehat{\gamma})=\frac{1}{N}\sum_{m=1}^{N}\big|u^{s}_{m}-u^{mes}_{m}
\big|^{2},
\end{equation}
where $u^{mes}_{m}$ and $u^{s}_{m}$ represent the measurements and the numerical solution of the scattered field at the $m$th observation points $(x_{m},h), x_m=\frac{2\pi m}{N},m =0,1, 2, \ldots, N$, respectively. According to the definition of $C(\widehat{\gamma})$, we have
\begin{equation}\label{PD3}
\frac{\partial C(\widehat{\gamma})}{\partial
\gamma_{j}}=\Re\bigg(\frac{2}{N}\sum_{m=1}^{N}\big(\overline{u^{s}_{m}(\widehat{\gamma})-u^{mes}_{m}
}\big)\frac{\partial\mathcal{P}(\widehat{\gamma})}{\partial
\gamma_{j}}\bigg),
\end{equation}
which implies that the gradient of the objective function $C(\widehat{\gamma})$ exists
$$
\nabla C(\widehat{\gamma}):=\bigg[\frac{\partial C(\widehat{\gamma})}{\partial
\gamma_{0}},\frac{\partial C(\widehat{\gamma})}{\partial
\gamma_{1}},\cdots,\frac{\partial C(\widehat{\gamma})}{\partial
\gamma_{2M}}\bigg]^{\top}.
$$

Next, we consider the following case of phaseless data. After defining the operator
$$
\mathcal{Q}(\widehat{\gamma}):=\mathcal{P}(\widehat{\gamma})\overline{\mathcal{P}(\widehat{\gamma})}
=|u^{s}_{\widehat{\gamma}}(x,h)|^{2},
$$
then, we have its derivative with respect to $\gamma_j$ is given by
\[
\frac{\partial\mathcal{Q}(\widehat{\gamma})}{\partial
\gamma_{j}} = 2\Re\bigg(\overline{u_{\widehat{\gamma}}^{s}(x,h)}\frac{\partial\mathcal{P}(\widehat{\gamma})}{\partial
\gamma_{j}}\bigg).
\]

In the case when the measurements of the scattered fields are the phaseless data. We can rewrite the discrete form of the objective function as
\begin{equation}\label{PLD1}
C(\widehat{\gamma})=\frac{1}{N}\sum_{m=1}^{N}\big(|u^{s}_{m}|^2-\big|u^{mes}_{m}
\big|^{2}\big)^{2},
\end{equation}
where $|u^{mes}_{m}|$ and $|u^{s}_{m}|$ represent the phaseless measurements and the modulus of numerical solution of the scattered field at the $m$th observation points $(x_{m},h), x_m=\frac{2\pi m}{N},m =0, 1, 2, \ldots, N$, respectively. From \eqref{PLD1}, we have
\begin{equation}\label{PLD2}
\frac{\partial C(\widehat{\gamma})}{\partial
\gamma_{j}}=\frac{2}{N}\sum_{m=1}^{N}\big(\big|u^{s}_{m}(\widehat{\gamma})\big|^{2}-\big|u^{mes}_{m}
\big|^{2}\big)\frac{\partial\mathcal{Q}(\widehat{\gamma})}{\partial
\gamma_{j}},
\end{equation}
then the gradient $\nabla C(\widehat{\gamma})$ is obtained via \eqref{PLD2}.

Thus, using the Broyden family's method\upcite{RF}, the {\bf{reconstruction algorithm}} is summarized as follows:


{\bf Step 1.}\ The finite element method, as detailed in \cite{ZMZ}, is employed for the numerical solution of the direct problem. Given an initial guess $\widehat{\gamma}^{(k)}$, in the case of phase data, leveraging Equations \eqref{OP} and \eqref{PD3}, the gradient of the cost function, $\nabla C(\widehat{\gamma}^{(k)})$ can be obtained using the central difference approximation. For phaseless data, the gradient $\nabla C(\widehat{\gamma}^{(k)})$ is calculated using Equation \eqref{PLD2}.

{\bf Step 2.}\ Choosing an accuracy $ \epsilon> 0$ and a symmetric positive definite matrix $B_k$, computing $ \overline{p}_k $
$$
\overline{p}_k =-B_k^{-1}\nabla C(\widehat{\gamma}^{(k)}).
$$

{\bf Step 3.}\ By using a line search to get a suitable stepsize $\alpha_k$ in this direction $\overline{p}_k $ such that
$$
C(\widehat{\gamma}^{(k)} +
\alpha_k\overline{p}_k)=\min_{\alpha\geq0}C(\widehat{\gamma}^{(k)} +
\alpha\overline{p}_k),
$$
then update
$$
\widehat{\gamma}^{(k+1)} = \widehat{\gamma}^{(k)} +
\alpha_k\overline{p}_k.
$$

{\bf Step 4.}\ Set
$$
\overline{s}_k=\alpha_k \overline{p}_k,\ \ \ \ \ \overline{y}_k =
{\nabla C(\widehat{\gamma}^{(k+1)}) - \nabla C(\widehat{\gamma}^{(k)})},\ \ \ \ \
$$
and
\begin{eqnarray*}
B_{k+1}=(1-\varphi)\bigg(B_k + \frac{\overline{y}_k
\overline{y}_k^{\mathrm{T}}}{\overline{y}_k^{\mathrm{T}}
\overline{s}_k} - \frac{B_k \overline{s}_k
\overline{s}_k^{\mathrm{T}} B_k }{\overline{s}_k^{\mathrm{T}} B_k
\overline{s}_k}\bigg)
+\varphi\bigg( \bigg(I-\frac{\overline{y}_k
\overline{s}_k^{\mathrm{T}}}{\overline{y}_k^{\mathrm{T}}
\overline{s}_k}\bigg) B_k \bigg(I-\frac{\overline{s}_k
\overline{y}_k^{\mathrm{T}}}{\overline{y}_k^{\mathrm{T}}
\overline{s}_k}\bigg) +\frac{\overline{y}_k
\overline{y}_k^{\mathrm{T}}}{\overline{y}_k^{\mathrm{T}}
\overline{s}_k}\bigg),
\end{eqnarray*}
where $\varphi \in [0,1]$ is a parameter, and $B_0 = I$ is an initial approximate
Hessian matrix. If
$$
\|\widehat{\gamma}^{(k+1)}-\widehat{\gamma}^{(k)}\|_2\leq\epsilon
$$ or
the number of iterations reaches the maximum $IT_{Max}$,
stop the algorithm. Otherwise,
letting $k=k+1$, and repeat Steps 1-4.

The above algorithm will generate
a sequence $\{\widehat{\gamma}^{(k)}\}_{k=1}^{K}$ of approximations to the minimum $\widehat{\gamma}^{*}$ of $C(\widehat{\gamma})$.
Thus, the reconstruction of the periodic structure is obtained
\[
f_{K}(x)= \gamma_{0}^{(K)}+\sum_{m=1}^{M}\big(\gamma_{2m-1}^{(K)}\cos(mx)
+\gamma_{2m}^{(K)}\sin(mx)\big).
\]

\section{Numerical results}\label{section: Numerical}

This section presents the numerical implementation and testing of the proposed illumination strategy and algorithm for reconstructing the grating profile from both near-field phase and phaseless measurements. 

Let $\vartheta=[\vartheta_{m}]_{N\times1}$ denote a random vector, we note that there exists a disturbance in data
such that the measurements $u_m^{mes}$ and $|u_{m,\delta}^{mes}|$ are replaced by, respectively,
\begin{equation}\label{MD1}
u_{m,\delta}^{mes}=u_m^{mes}\bigg(1
+\delta\frac{\vartheta_m}{\|\vartheta\|_{2}}\bigg),
\quad\quad m = 1, 2, \ldots, N,
\end{equation}
and
\begin{equation}\label{MD1}
|u_{m,\delta}^{mes}|^{2}=|u_m^{mes}|^{2}
+\delta\bigg(\frac{\|u^{mes}\|^{2}_{2}}{\|\vartheta\|^{2}_{2}}\bigg)\vartheta_m,
\quad\quad m = 1, 2, \ldots, N,
\end{equation}
where
$
\|u^{mes}\|_{2}:=\bigg(\sum\limits_{m=1}^{N}|u_m^{mes}|^{2}\bigg)^{\frac{1}{2}},
 \|\vartheta\|_{2}:=\bigg(\sum\limits_{m=1}^{N}|\vartheta_{m}|^{2}\bigg)^{\frac{1}{2}}
$
and $\delta$ is the noise level.

To illustrate the application of the superposition incident wave strategy, we first consider an exemplary case.

{\bf{Example 4.1.}} Given the superposition incident field $u^{\rm I}$ satisfying \eqref{IF}, for $x\in (0,2\pi)$, considering the grating structure as
\begin{align*}
f(x)&=0.2\cos(x)+0.1\sin(x)-0.2\cos(2x)-0.1\sin(2x)+0.1\cos(3x)-0.2\sin(3x)\nonumber\\
    &\quad-0.1\cos(4x)-0.2\sin(4x)+0.2\cos(5x)-0.4\sin(5x),
\end{align*}
which implies that $\widehat{\gamma}^{(*)}=[0,0.2,0.1,-0.2,-0.1,0.1,-0.2,-0.1,-0.2,0.2,-0.4]^{\top}$ and
$M=5$. 

We consider the exemplary case with  $N=256$, initial guess $\widehat{\gamma}^{(0)}=[0]^{\top}_{11\times1}$, noise level $\epsilon=10^{-3}$ and maximum iteration count $IT_{Max}=50$. Algorithm 3.1 is employed for reconstructing the aforementioned grating, with the parameter of measuring position set to  $h=1.0$. Figure \ref{PD1} presents the reconstructed surfaces (dashed lines) obtained for $\kappa=1$, and $\kappa=2$, respectively. As evident from the figure, super-resolution imaging is achieved, and a larger value of $\kappa$ (Figure \ref{PD1}(b)) leads to a more accurate reconstruction compared to the case with a smaller $\kappa$ (Figure \ref{PD1}(a)).

To further explore the influence of the measuring position parameter, $h$, on the reconstructed grating profile, Figure \ref{PD2} presents the results obtained using $\kappa=3$ and a noise level of $\delta=10\%$. The figures depict reconstructions for varying values of $h=0.8$ (Figure \ref{PD2}(a)), $1.0$ (Figure \ref{PD2}(b)), $1.2$ (Figure \ref{PD2}(c)), and $1.4$ (Figure \ref{PD2}(d)). As observed from the figures, a smaller value of $h$ (Figure \ref{PD2}(a)) leads to a more accurate reconstruction compared to larger values (Figures \ref{PD2}(b)-(d)).

\begin{figure}[http]
\centering
\subfigure[]{
\includegraphics[width=7.3cm]{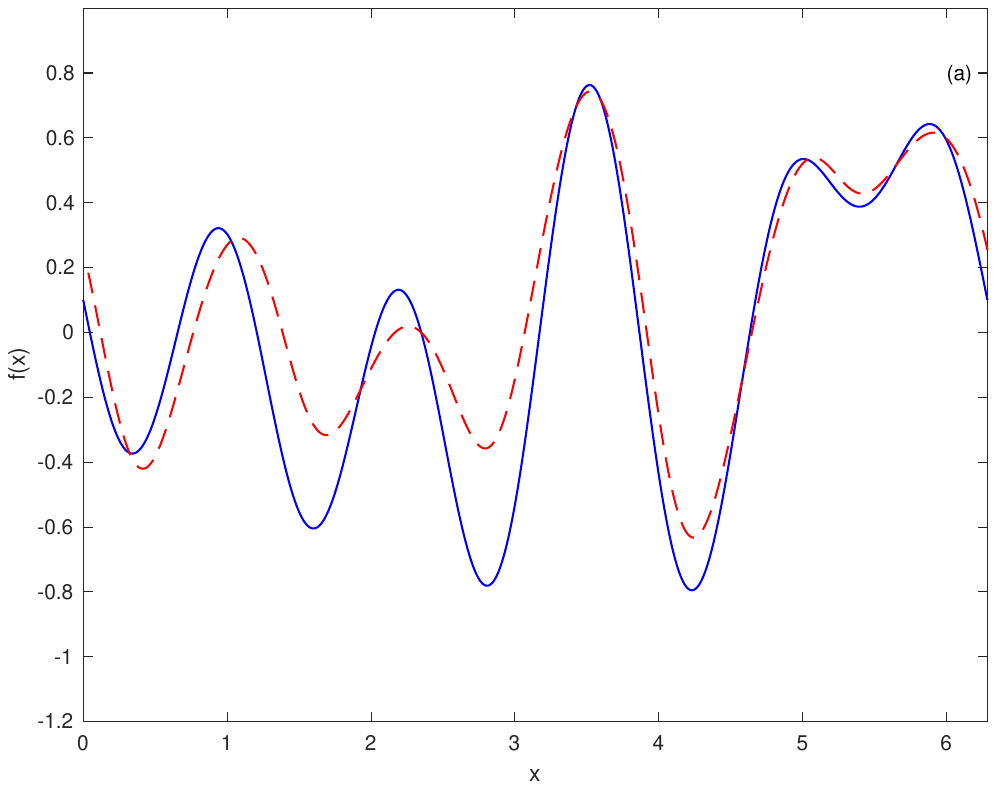}
}
\quad
\subfigure[]{
\includegraphics[width=7.3cm]{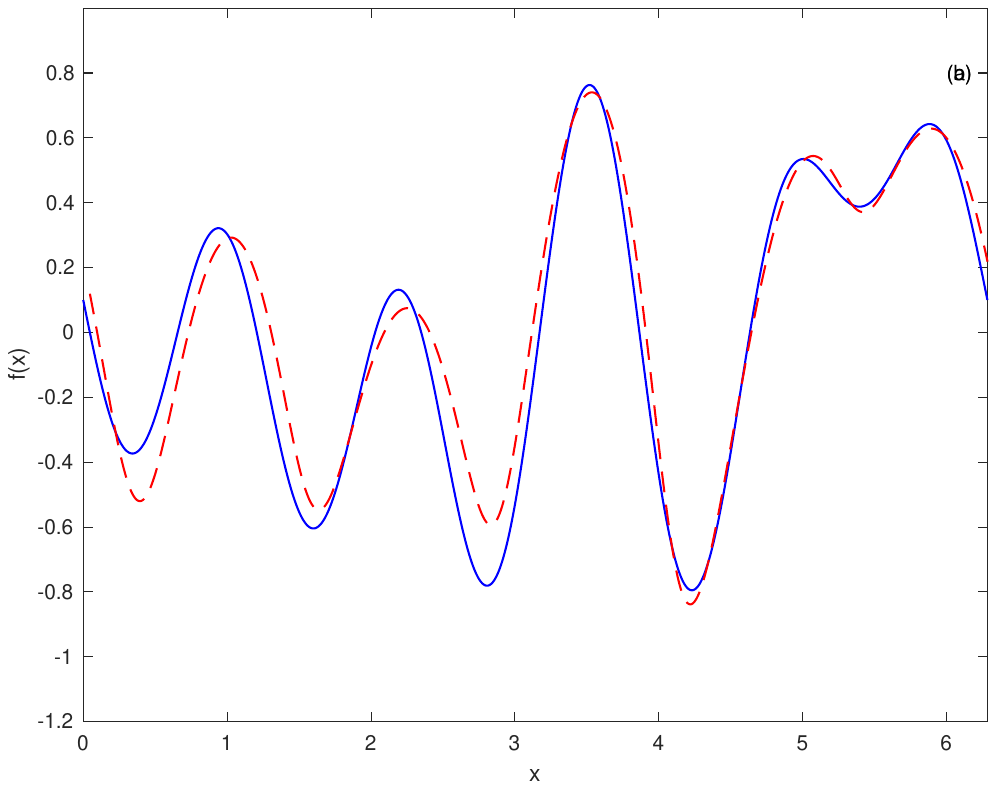}
}
\caption{The numerical results (dashed lines) and the exact gratings (solid lines) are plotted for the superposition incident field $u^{\rm I}$ with $h=1.0,\delta=5\%$. (a) $\kappa=1$; (b) $\kappa=2$.}
\label{PD1}
\end{figure}

\begin{figure}[htbp]
\centering
\subfigure[]{
\includegraphics[width=7.3cm]{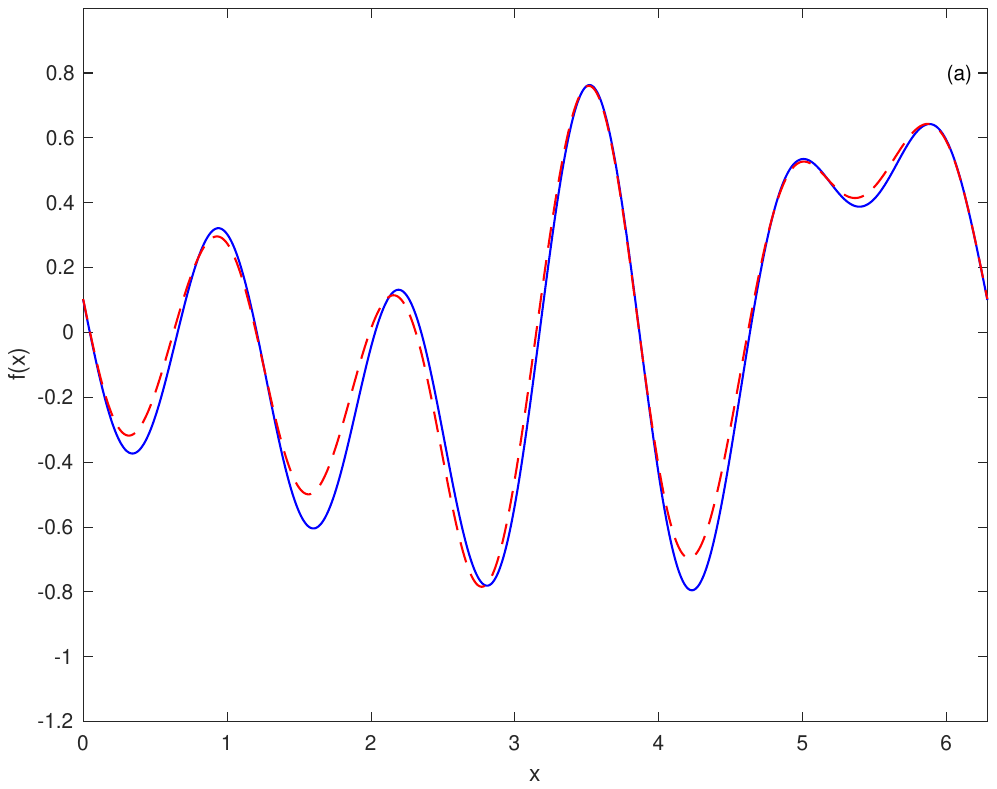}
}
\quad
\subfigure[]{
\includegraphics[width=7.3cm]{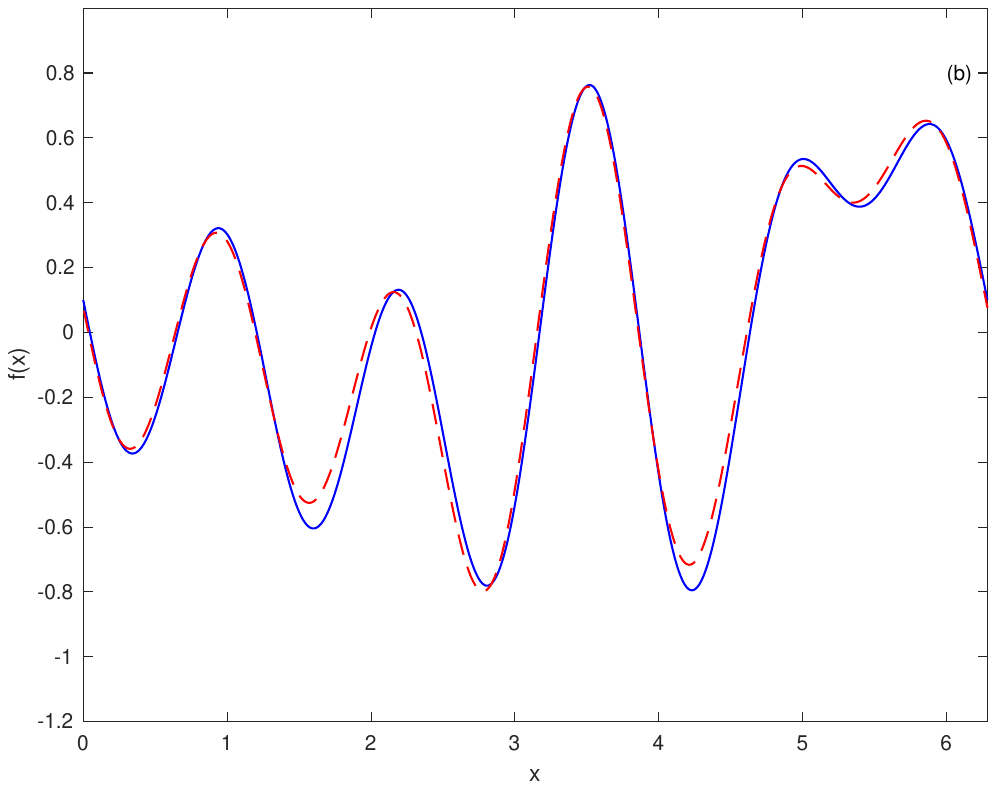}
}
\quad
\subfigure[]{
\includegraphics[width=7.3cm]{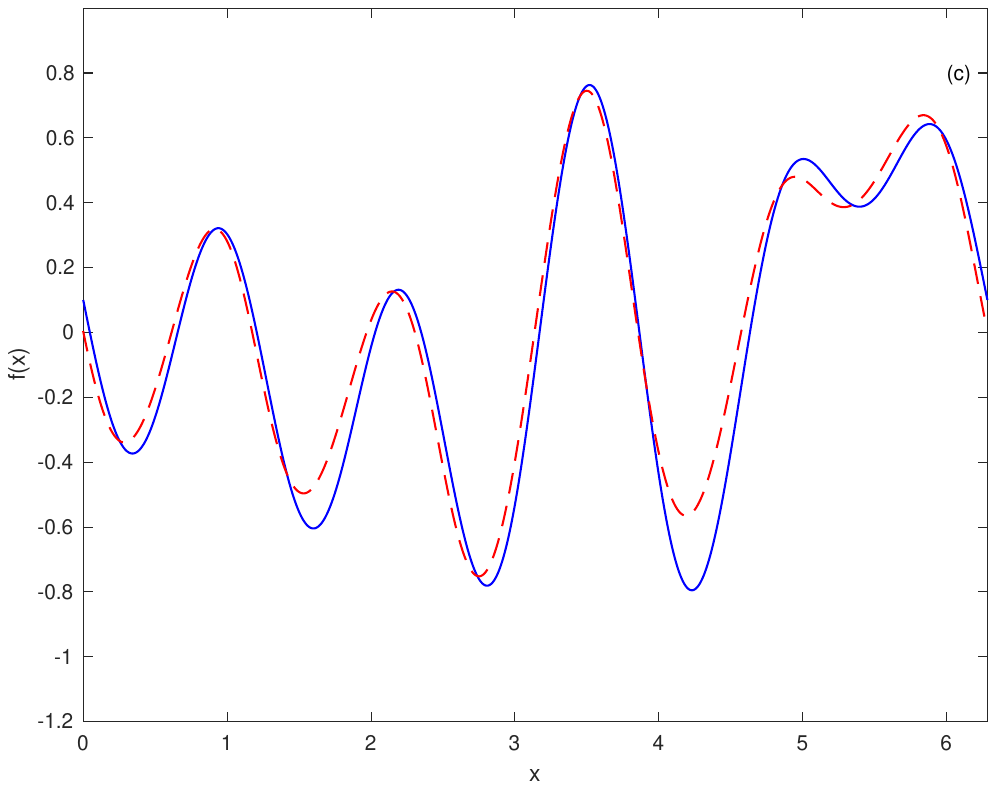}
}
\quad
\subfigure[]{
\includegraphics[width=7.3cm]{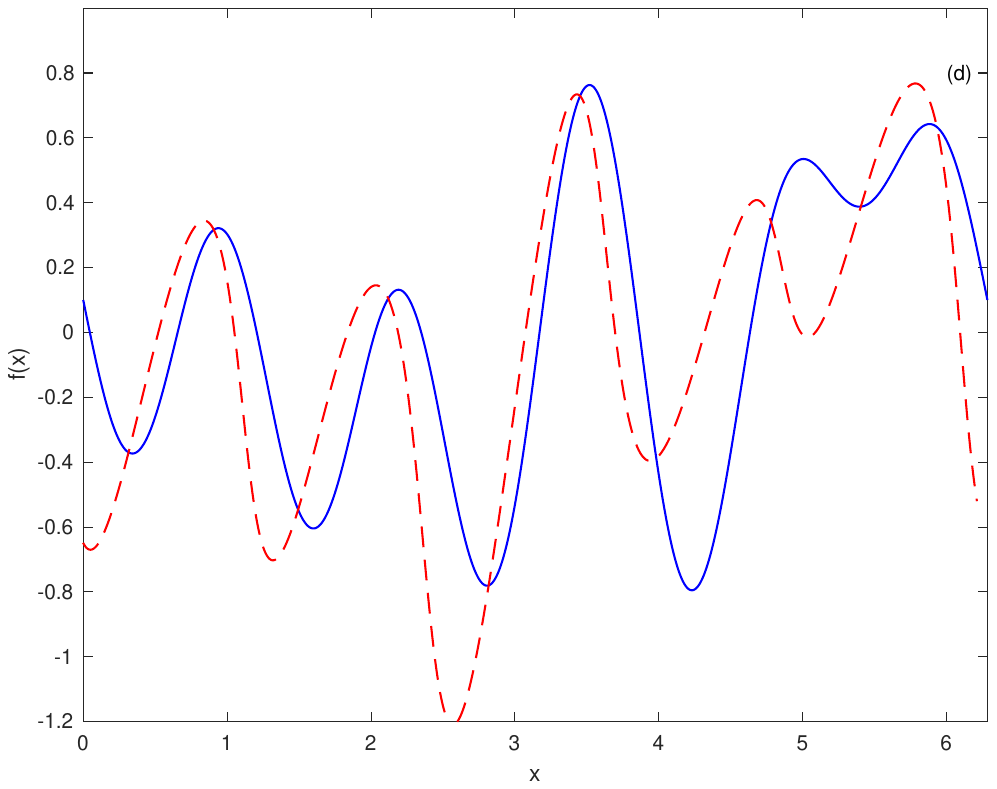}
}
\caption{The numerical results (dashed lines) and the exact
gratings (solid lines) are plotted for the incident field $u^{\rm I}$ with $\kappa=3, \delta=10\%$. (a) $h=0.8$; (b) $h=1.0$; (c) $h=1.2$; (d) $h=1.4$.}
\label{PD2}
\end{figure}

To further investigate the dependence of the reconstruction results on the incident field, we present an additional numerical example utilizing phaseless measurements.

{\bf{Example 4.2.}} For $x\in (0,2\pi)$, considering the grating structure as
\begin{align*}
f(x)=0.2\sin(x)+0.1\cos(2 x)+0.3\sin(3 x),
\end{align*}
which implies that $\widehat{\gamma}^{(*)}=[0,0,0.2,0.1,0,0,0.3]^{\top}$ and
$M=3$. 

\begin{figure}[htbp]
\centering
\subfigure[]{
\includegraphics[width=7.3cm]{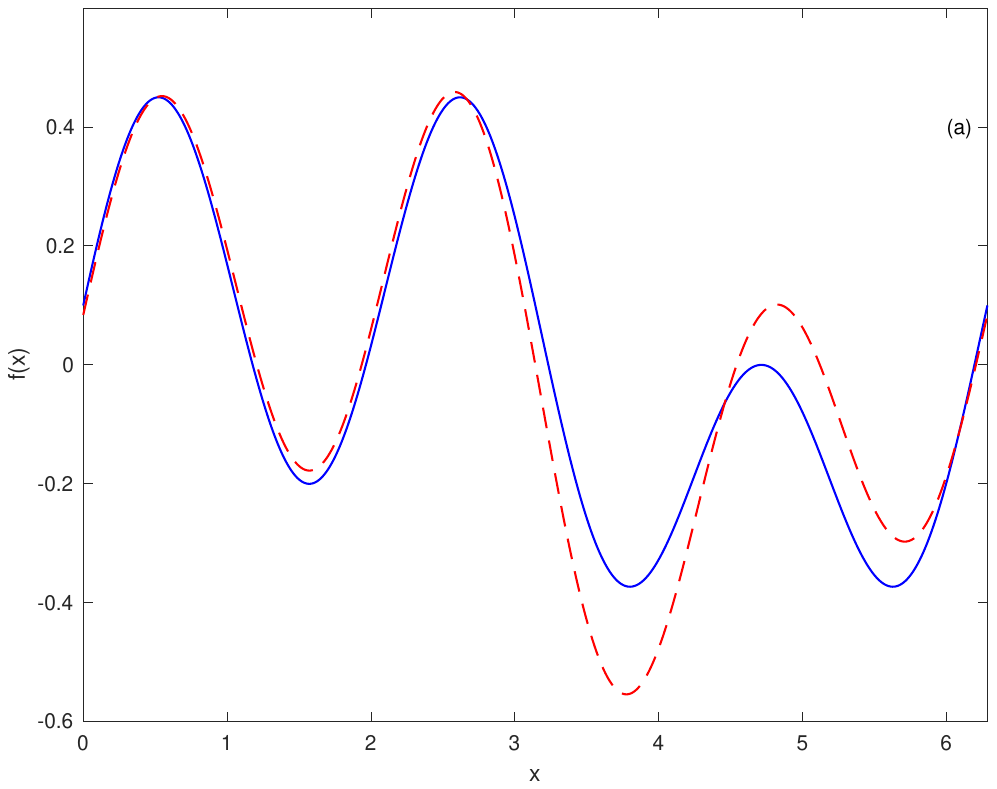}
}
\quad
\subfigure[]{
\includegraphics[width=7.3cm]{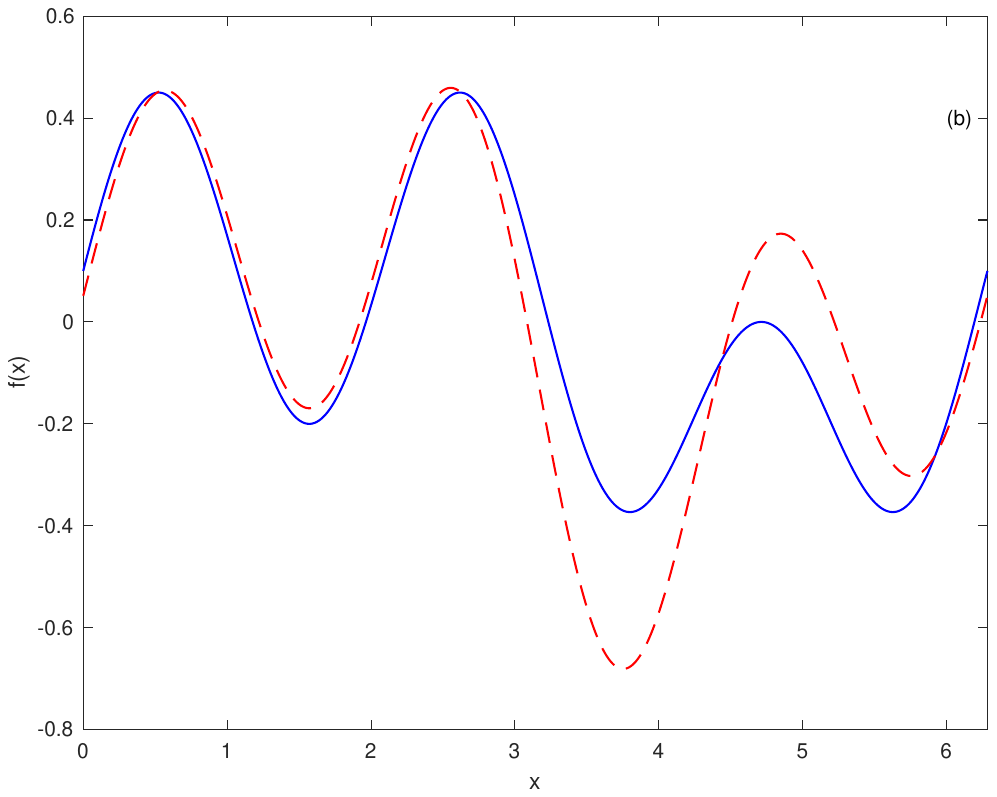}
}
\caption{The numerical results (dashed lines) and the exact
gratings (solid lines) are plotted for the superposition incident field $u^{\rm I}$ with $\kappa=1$. (a) $\delta=5\%$; (b) $\delta=10\%$.}
\label{abc}
\end{figure}

\begin{figure}[htbp]
\centering
\subfigure[]{
\includegraphics[width=7.3cm]{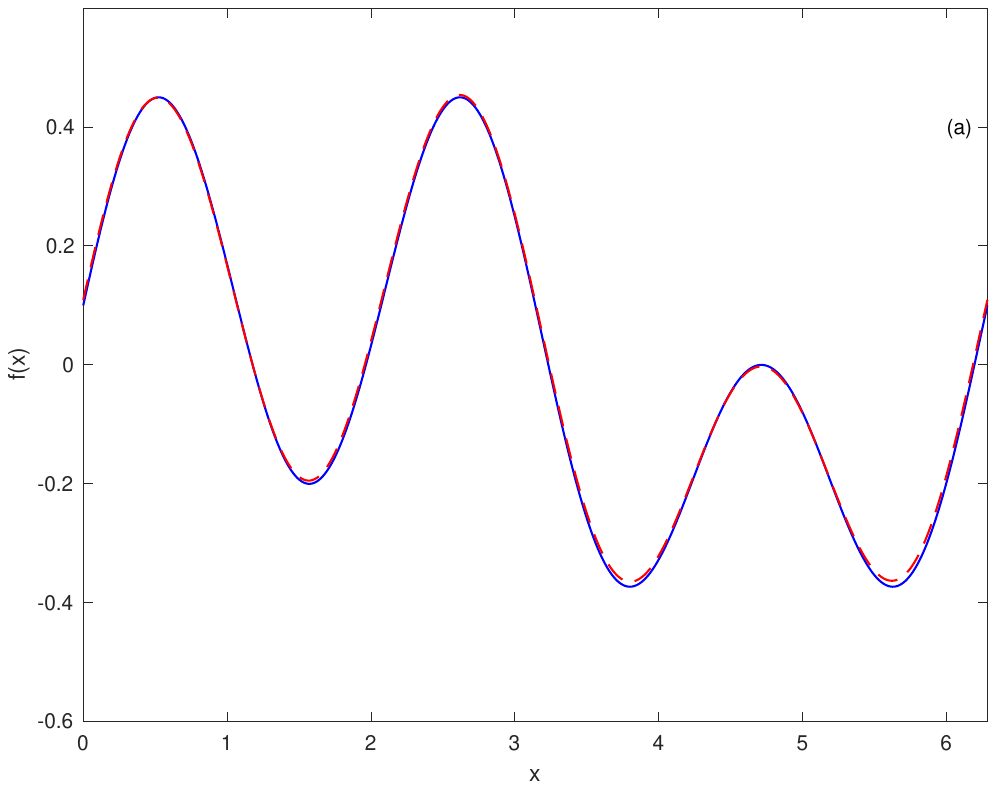}
}
\quad
\subfigure[]{
\includegraphics[width=7.3cm]{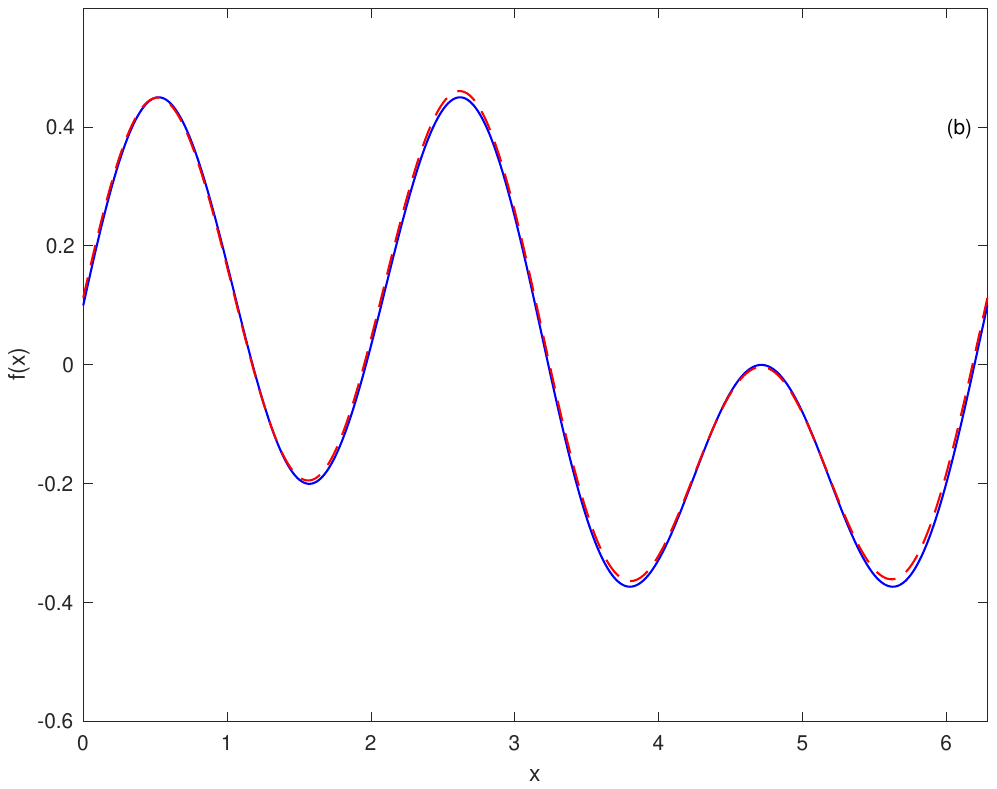}
}
\caption{The numerical results (dashed lines) and the exact
gratings (solid lines) are plotted for the superposition incident field $u^{\rm I}$ with $\kappa=2$. (a) $\delta=5\%$; (b) $\delta=10\%$.}
\label{abck2}
\end{figure}

We set $h=0.8, N=256,\widehat{\gamma}^{(0)}=[0]^{\top}_{7\times1}, \epsilon=10^{-3}$ and
$IT_{Max}=50$. Initially, we investigate reconstructions utilizing the superposition incident field, $u^{\rm I}$. With a fixed value of $\kappa=1$, Figures \ref{abc}(a) and \ref{abc}(b) depict the reconstructed surfaces (dashed lines) obtained for noise levels of  $\delta=5\%$ and $\delta=10\%$, respectively. As observed in these figures, the algorithm demonstrates stable reconstruction of the exact surface shape, even with $10\%$ noise, exhibiting performance comparable to the $5\%$ noise case. Notably, in the context of the periodic structure, the distance between peaks and troughs is approximately $\lambda/6$, indicating that the algorithm achieves super-resolution imaging. Indeed, Figure \ref{abck2} demonstrates the better reconstruction results for a larger value of $\kappa=2$.

\begin{figure}[htbp]
\centering
\subfigure[]{
\includegraphics[width=7.3cm]{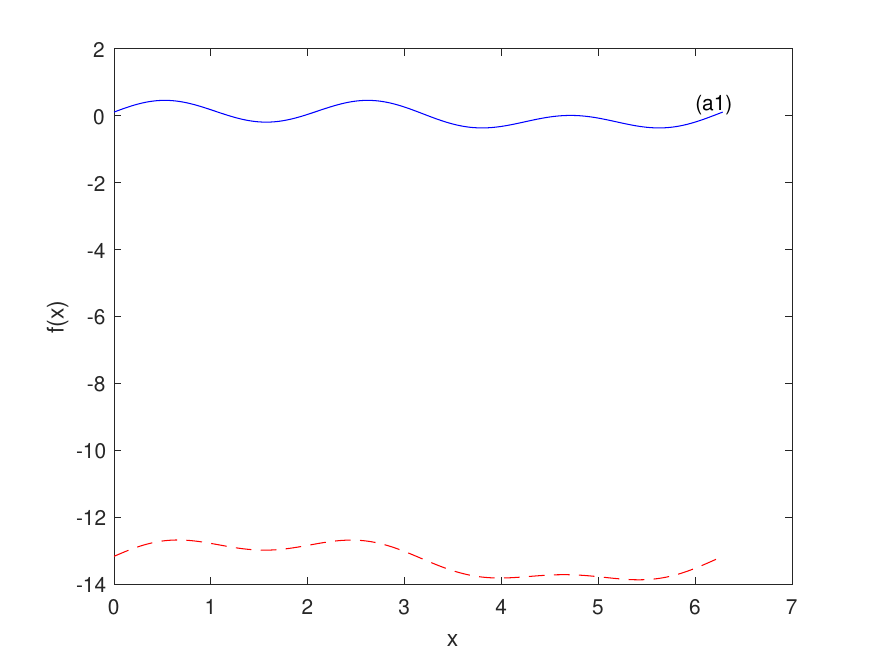}
}
\quad
\subfigure[]{
\includegraphics[width=7.3cm]{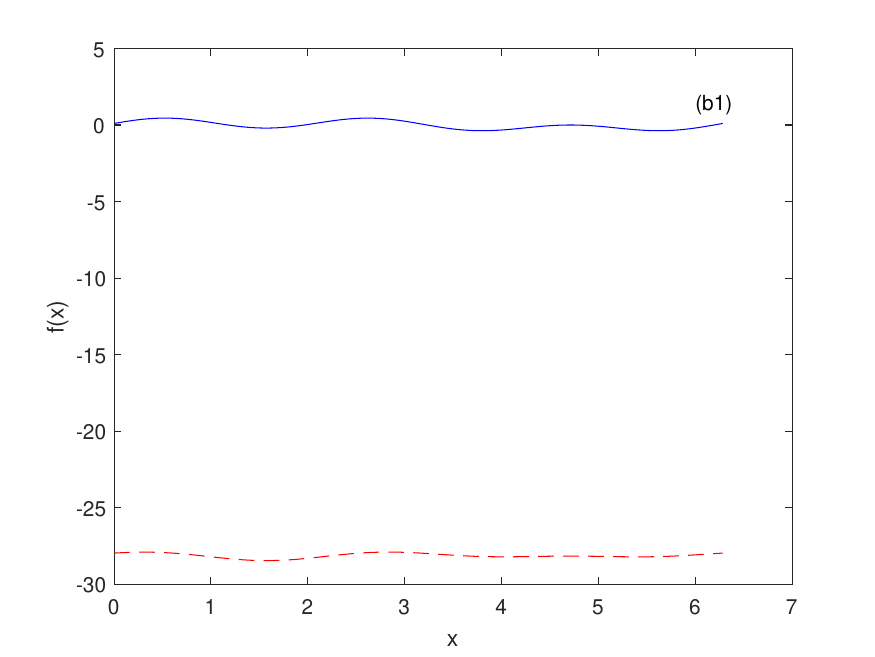}
}
\caption{The numerical results (dashed lines) and the exact
gratings (solid lines) are plotted for the incident field $u^{\rm inc}$ with $\kappa=2$. (a) $\delta=0\%$; (b) $\delta=5\%$.}
\label{abci}
\end{figure}

Figure \ref{abci} presents the reconstruction results obtained solely with a plane incident field $u^{\rm inc}$, for noise levels of $\delta=0\%$ and $5\%$, with a fixed value of $\kappa=2$. As evident from the figure, no reasonable reconstruction results were obtained in this scenario, indicating the limitations of this approach.

\begin{figure}[htbp]
\centering
\subfigure[]{
\includegraphics[width=7.3cm]{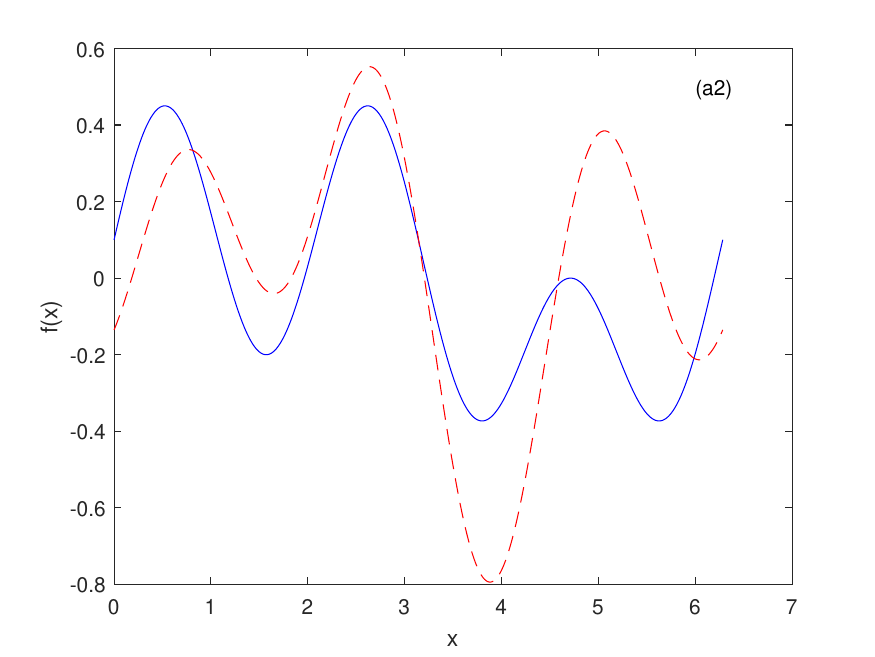}
}
\quad
\subfigure[]{
\includegraphics[width=7.3cm]{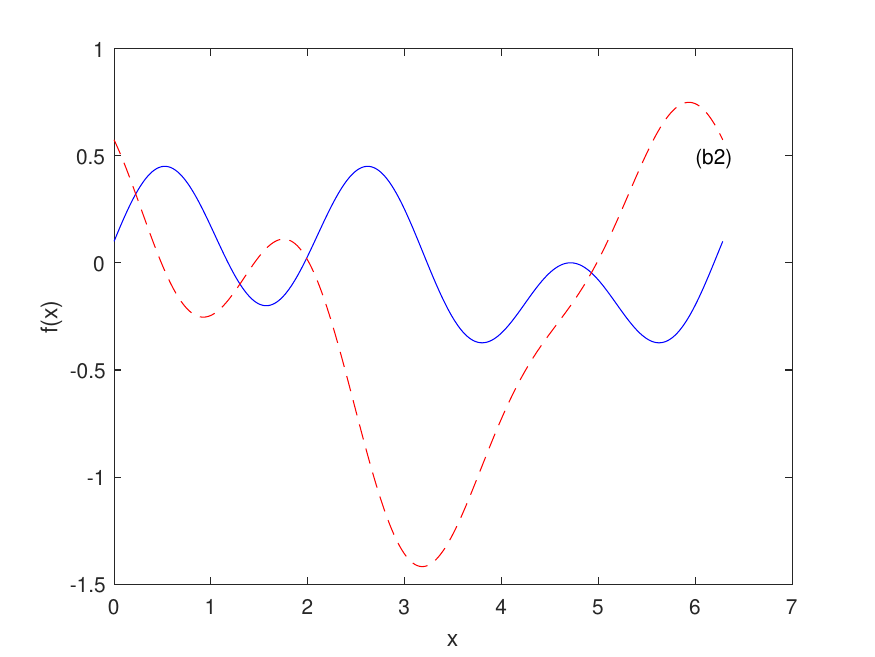}
}
\caption{The numerical results (dashed lines) and the exact
gratings (solid lines) are plotted for the incident field $\widetilde{u^{\rm inc}}$ with $\kappa=2$. (a) $\delta=0\%$; (b) $\delta=5\%$.}
\label{abci1}
\end{figure}

Figure \ref{abci1} depicts the reconstructions obtained solely using the incident field $\widetilde{u^{\rm inc}}$, under the same conditions (fixed $\kappa=2$ and $\delta=0\%$ and $5\%$). Comparing Figures \ref{abci1}(a) and \ref{abci1}(b), it is evident that stable reconstruction of the exact surface shape is not archived with $5\%$ noise using only $\widetilde{u^{\rm inc}}$, exhibiting a deterioration in performance compared to the no-noise case.

The presented numerical results corroborate the potential advantages of employing superimposed incident fields, as discussed in Chapter 3.

\section{Conclusions}\label{section: Con}

This paper investigates the application of a superposition illumination strategy to achieve super-resolution reconstruction of periodic structures from phase or phaseless near-field measurements. Our work presents two key innovations:

Theoretical analysis of the advantages of superposition illumination strategy: Utilizing the superposition illumination strategy and the transparent boundary condition (TBC), we establish a well-posed direct scattering problem. We demonstrate that, for a fixed incident field $u^{\rm I}$ in a specific direction, the corresponding scattered fields can offer distinct advantages for numerical reconstructions. Furthermore, within a specific wavenumber interval, analysis of the Laplacian operator eigenvalues reveals that the scattered fields are sufficient to uniquely identify the grating structure.

Effective reconstruction algorithm and numerical results: We present an efficient algorithm for reconstructing the periodic structure from near-field measurements. Numerical results elucidate the influence of incident wave fields on grating profile inversion. Notably, the superposition incidence strategy demonstrably improves the conditioning of the reconstruction problem, enhancing the signal-to-noise ratio of measurements while simultaneously resolving the issue of up-down translational invariance and facilitating super-resolution imaging compared to using a single incident wave.





\bibliographystyle{elsarticle-num}


\end{document}